\documentclass[9pt,twocolumn,twoside]{osajnl}
\journal{ol}
\setboolean{shortarticle}{true}
\usepackage{graphicx}
\usepackage{txfonts}
\usepackage[english]{babel}
\newlength{\figsize}
\newlength{\subfigsize}
\begin{document}

\title{Optical Kerr nonlinearity of arrays of all-dielectric high index nanodisks in the vicinity of the anapole state}
\author{Andrey V. Panov}

 \affil[1]{
Institute of Automation and Control Processes,
Far Eastern Branch of Russian Academy of Sciences,
5, Radio st., Vladivostok, 690041, Russia}
\affil[*]{Corresponding author: andrej.panov@gmail.com}
\dates{Received 4 March 2020; revised 16 April 2020; accepted 29 April 2020; published 28 May 2020}
\setboolean{displaycopyright}{true}
\setcounter{page}{3071}
\renewcommand*{\journalname}{Vol. 45, No. 11 (2020) // Optics Letters}
\doi{\url{https://doi.org/10.1364/OL.391991}}

\begin{abstract}
The nonlinear optical properties of the high index nanoparticles are boosted at the anapole state.
Researchers intensively study this phenomenon as promising for various applications.
In this work, the nonlinear optical Kerr effect of disordered  and square lattice metasurfaces of GaP nanodisks is investigated as a function of the disk size in the vicinity of the anapole state at the wavelength of 532 nm. 
When the sizes of the nanodisks are close to the anapole state, the effective second order refractive index of the metasurface increases exponentially. 
On approaching the anapole state, the sign of the effective second order refractive index is inverted.
The absolute value of the effective nonlinear Kerr coefficient of the square lattice metasurface is higher than that of the
disordered array of nanodisks.
The absolute value of the effective second order refractive index in proximity to the anapole state is an order of magnitude higher than that at non-anapole resonances of the disordered metasurfaces consisting of the nanodisks or spheres.
\end{abstract}
\maketitle

In recent years, considerable attention of researchers has been given to the study of the anapole states of high-index all-dielectric nanostructures. 
The term anapole means the radiationless state of the nanoparticle. 
This state arises from the destructive interference of the radiation fields of electric dipole resonance with an oscillating toroidal dipole moment leading to vanishing scattering.
The toroidal multipoles are the combinations of the higher order terms of an expansion of the electric multipole modes with respect to the electromagnetic size of the source \cite{Fernandez17}.
The optical anapole modes result in enormous confinement of electric energy inside the scatterer.
The usage of anapole states has been proposed for designing the metamaterials or metasurfaces with anapole-induced transparency \cite{Fedotov13,Ospanova18}, invisibility \cite{Liu15} and enhanced nonlinear effects. 

Baranov et al.\ \cite{Baranov18} demonstrated two orders of magnitude enhancement of Raman scattering intensity by Si disk array at the anapole state compared to unpatterned Si film. 
At the same time, they revealed that the elastic scattering by Si nanodisks is suppressed due to destructive interference between the dipole electric and toroidal modes.
Grinblat et al.\ showed that third-harmonic conversion efficiency of a germanium nanodisk at the anapole state at an excitation wavelength of 1650 nm is four orders of magnitude greater than in the case of an unstructured Ge film \cite{Grinblat16}. 
Also, the anapole states at nanodisks or nanodimers of non-centrosymmetric materials (e.g. AlGaAs) can be utilized for enhanced second-harmonic generation \cite{Rocco18,Timofeeva18}. 
Timofeeva et al.\ \cite{Timofeeva18} observed that the conversion efficiency of the second-harmonic generation of the AlGaAs free-standing disks is around $5\times10^3$ times larger than that of the continuous layer of this material.
Grinblat et al.\ \cite{Grinblat18} demonstrated that in the vicinity of the first-order anapole mode single Si nanodisks covered with a layer of Au can be used for ultrafast all-optical switching due to the nonlinear optical Kerr effect.

Recently, it was proposed a computational method for retrieving the effective Kerr nonlinearity of nanocomposite media~\cite{Panov18}. 
This method is based on three-dimensional finite-difference time-domain (FDTD) simulations of light propagation.
Previously, by three-dimensional FDTD modeling, the optical Kerr 
effect of disordered nanocomposites consisting of high index Mie-resonant spheres 
was studied \cite{Panov19}. It was shown that the intensity dependent effective 
refractive index grows on the orders of magnitude in approaching Mie 
resonances. In vicinity of the resonances, the sign of the second order 
refractive index is inverted.


\begin{figure}
{\centering\includegraphics[width=\figsize]{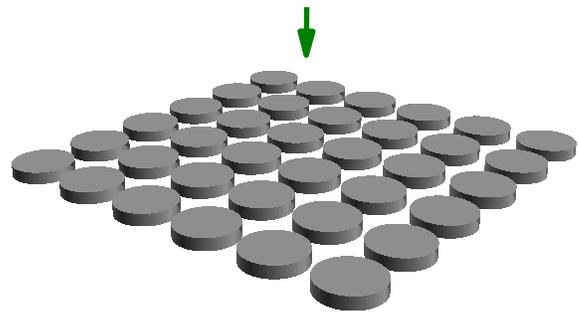}\par} 
\caption{\label{nldisks} (Color online) 
Schematic of the simulated metasurface consisting of the high index nanodisks. 
The Gaussian beam is incident normally on the metasurface. }
\end{figure} 

In this Letter, the optical Kerr effect of the random and ordered
metasurfaces consisting of GaP nanodisks is investigated in the vicinity of the anapole state. 
This is done by means of three-dimensional finite-difference time domain simulations.
In this method, the phase change for different intensities of incident Gaussian beam is calculated after the transmission through the sample under investigation. 
A schematic of the metasurface is depicted in Fig.~\ref{nldisks}.
The dependence of the phase change on the incident light intensity $I$ gives the estimate of the nonlinear refractive index as defined by
\[
 n=n_0+n_2 I,
\]
where $n_0$ is the linear refractive index, $n_2$ is the second-order nonlinear refractive index.
The computations of the nonlinear index of refraction are performed at several locations on the transmitted beam axis allowing one to estimate the mean value and the standard deviation of  $n_2$.
Detailed description of the calculation procedure of the nonlinear refractive index is given in Ref.~\cite{Panov18}.

For modeling in this work, gallium phosphide (GaP) was selected as it
has high third-order nonlinearity and is transparent in visible range.
At $\lambda=532$ nm, the linear refractive index of GaP nanodisks $n_{0\,\mathrm{in}}$ was equal to 3.49 \cite{Aspnes83} and the extinction coefficient was neglected as its magnitude is small (0.0026).
The value of second-order nonlinear refractive index $n_{2\,\mathrm{in}}$ was set to $6.5\times10^{-17}$~m$^2$/W corresponding to  the measurements of the third-order optical susceptibility in the visible range $\chi^{(3)}_{\mathrm{in}}\approx 2 \times 10^{-10}$~esu \cite{Kuhl85}.
The size of the FDTD computational domain was $2\times 2\times 15$~$\mu$m, the space resolution of the simulations was 2.5~nm. 
The examined samples were disordered monolayers or square lattice comprising the identical GaP nanodisks.
The lattice constant was 310 nm with 36 disks on the $2\times 2$~$\mu$m$^2$ area.
The number of disks randomly arranged on the $2\times 2$~$\mu$m$^2$ area was 30 or 17.
The nanodisks did not overlap or touch each other.
The reduced number of particles (17) was used for the large disk sizes in order to arrange them on the fixed area.
The radii of the disks were varied and the thickness of the disks and their monolayers $h$ was always 50 nm as before the anapole states were observed for flat cylinders \cite{Miroshnichenko15}.
The Gaussian beam linearly polarized along $y$ direction with the beam radius at the beam waist of 5.5~nm falls perpendicularly on the disk monolayer.
The nanodisks were surrounded by vacuum.

\begin{figure}[tbh]
{\centering
\includegraphics[width=\figsize]{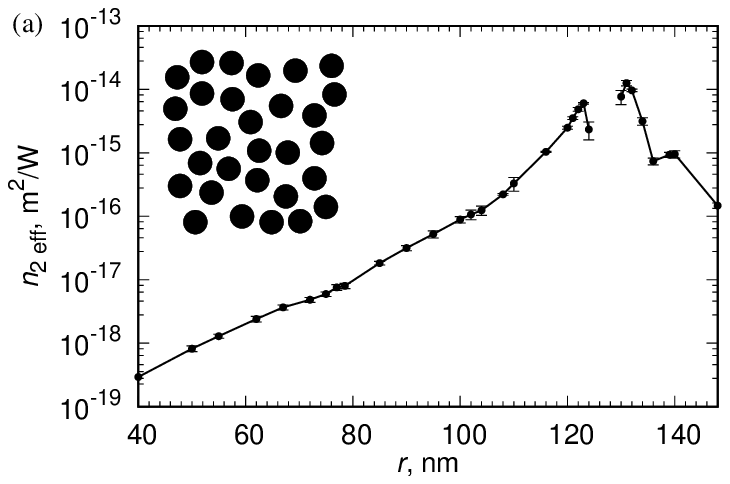}
\includegraphics[width=\figsize]{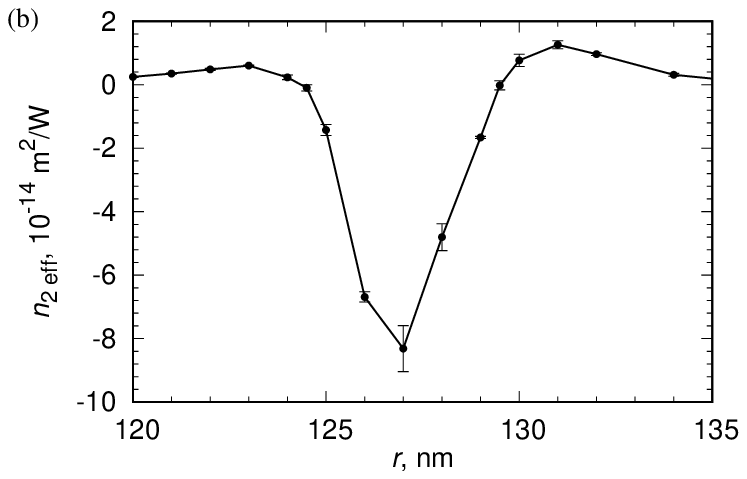}
\par} 
\caption{\label{n2_r_GaP_30_randcyl}
The effective second-order refractive index $n_{2\,\mathrm{eff}}$ of the disordered metasurfaces as a function of the GaP disk radius~$r$ at $\lambda=532$~nm for 30 particles arranged on the $2\times 2$~$\mu$m$^2$ area. In graph (a), the negative values of $n_{2\,\mathrm{eff}}$ are not displayed. Graph (b) shows the same data in the vicinity of the anapole state. The error bars show the standard deviations. The inset in (a) is a sketch of the nanostructure at the anapole state.}
\end{figure} 

\begin{figure}[tb]
{\centering
\includegraphics[width=\figsize]{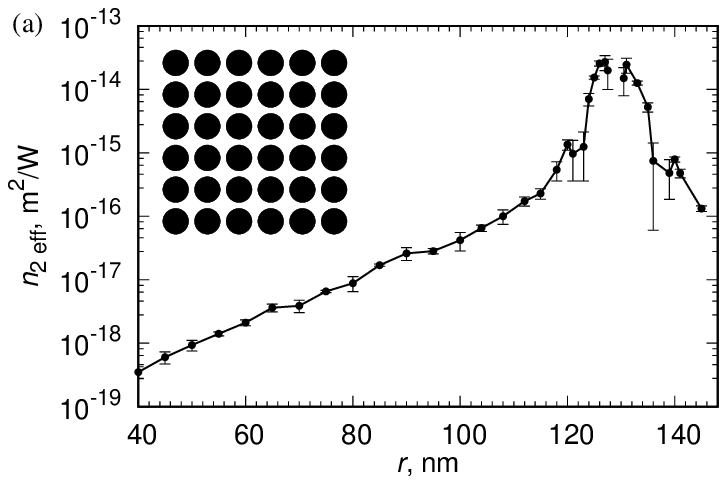}
\includegraphics[width=\figsize]{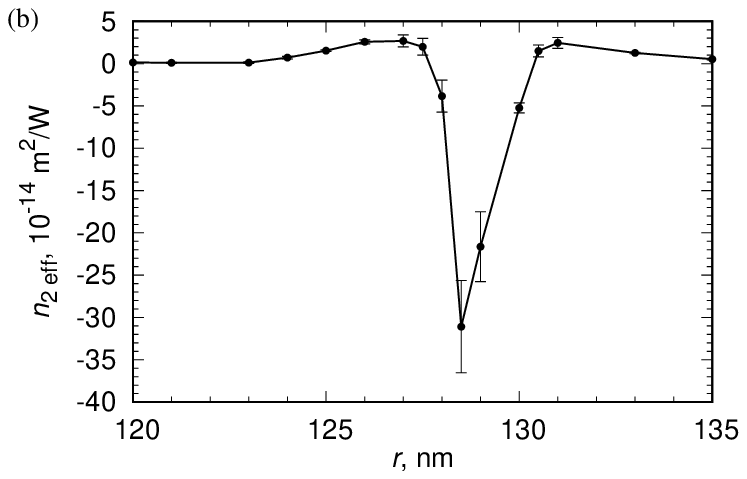}
\par} 
\caption{\label{n2_r_GaP_36_lattcyl}
The effective second-order refractive index $n_{2\,\mathrm{eff}}$ of the square lattices of the GaP disks with the lattice parameter of 310~nm as a function of radius~$r$ at $\lambda=532$~nm for 36 particles arranged on the $2\times 2$~$\mu$m$^2$ area. In graph (a), the negative values of $n_{2\,\mathrm{eff}}$ are not displayed. Graph (b) shows the same data in the vicinity of the anapole state. The error bars show the standard deviations. The inset in (a) is a sketch of the nanostructure at the anapole state.}
\end{figure} 

\begin{figure}[tb]
{\centering
\includegraphics[width=\figsize]{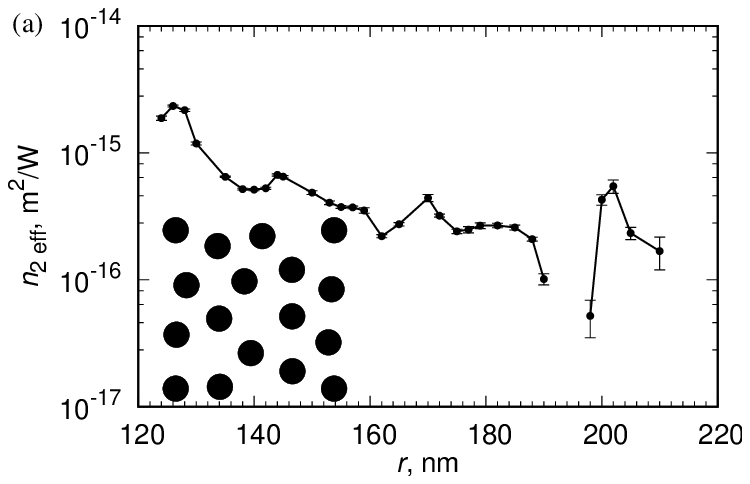}
\includegraphics[width=\figsize]{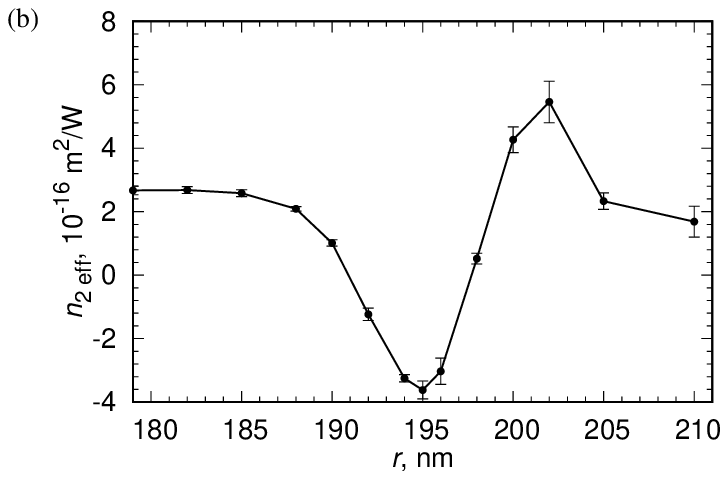}
\par} 
\caption{\label{n2_r_GaP_17_randcyl}
The effective second-order refractive index $n_{2\,\mathrm{eff}}$ of the disordered metasurfaces as a function of the GaP disk radius~$r$ at $\lambda=532$~nm for 17 particles arranged on the $2\times 2$~$\mu$m$^2$ area. In graph (a), the negative values of $n_{2\,\mathrm{eff}}$ are not displayed. Graph (b) shows the same data in the vicinity of a resonance. The error bars show the standard deviations. The inset in (a) is a sketch of the nanostructure at the anapole state.}
\end{figure} 

The dependencies of the effective second-order nonlinear refractive index of the metasurfaces with disordered arrangements or square lattices of the nanodisks on their sizes at $\lambda=532$~nm are depicted in Figs.~\ref{n2_r_GaP_30_randcyl}, \ref{n2_r_GaP_36_lattcyl}, \ref{n2_r_GaP_17_randcyl}. 
The data are plotted on a logarithmic scale for positive values of $n_{2\,\mathrm{eff}}$. 
At resonances, the effective second-order nonlinear refractive index can have a negative sign.
The vicinity of the resonance with negative values of $n_{2\,\mathrm{eff}}$ is shown separately on a linear scale in Figs~\ref{n2_r_GaP_30_randcyl}, \ref{n2_r_GaP_36_lattcyl}, \ref{n2_r_GaP_17_randcyl}.
Until the radii of about 120 nm, the effective second-order nonlinear refractive index of the metasurfaces nearly exponentially grows (Figs.~\ref{n2_r_GaP_30_randcyl}, \ref{n2_r_GaP_36_lattcyl}).
About $r=125$~nm, $n_{2\,\mathrm{eff}}$ abruptly drops to negative values. 
Then, the effective second-order nonlinear refractive index  becomes positive again.
The similar behavior of $n_{2\,\mathrm{eff}}$ of metasurfaces comprising GaP nanospheres at the Mie resonances was observed in Ref.~\cite{Panov19}. 
The size of this resonance for square lattice (Fig.~\ref{n2_r_GaP_36_lattcyl}) is slightly higher than for the disordered arrangement (Fig.~\ref{n2_r_GaP_30_randcyl}).
This is a typical behavior for densely packed nanoparticles whose resonance sizes are shifted to larger values.
In contrast, for the sparse metasurface the negative values of $n_{2\,\mathrm{eff}}$ are observed at $r<124$~nm (Fig.~\ref{n2_r_GaP_17_randcyl}).
At this resonance, the absolute values of $n_{2\,\mathrm{eff}}$ for the square lattice are several times higher than those of the disordered metasurface.

After that, the effective second-order nonlinear refractive index of the disordered metasurface decreases until $r=160$~nm (Fig.~\ref{n2_r_GaP_17_randcyl}).
About $r=195$~nm, there is another resonance with negative sign of $n_{2\,\mathrm{eff}}$.

\begin{figure}[tb]
{\centering
\includegraphics[width=\subfigsize]{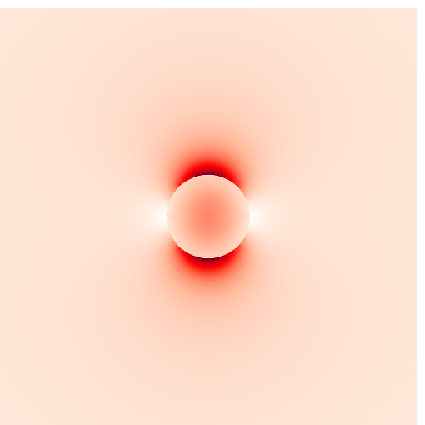}
\includegraphics[width=\subfigsize]{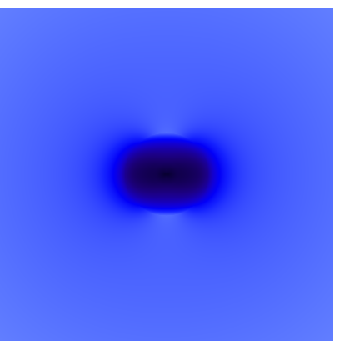}\\
{\centering
$r=50$~nm
\par}
\includegraphics[width=\subfigsize]{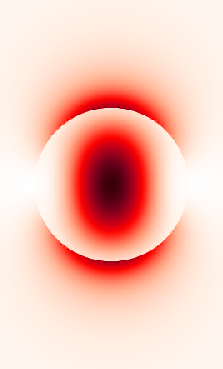}
\includegraphics[width=\subfigsize]{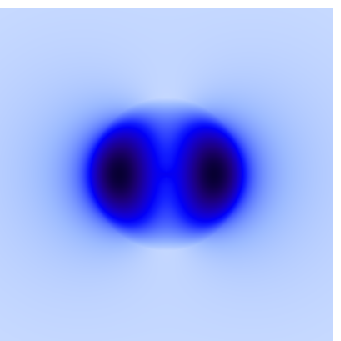}\\
{\centering
$r=92$~nm
\par}
\includegraphics[width=\subfigsize]{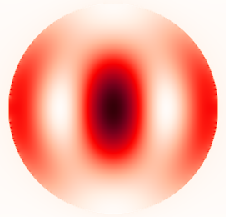}
\includegraphics[width=\subfigsize]{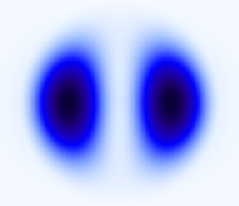}\\
{\centering
$r=126$~nm
\par}
\includegraphics[width=\subfigsize]{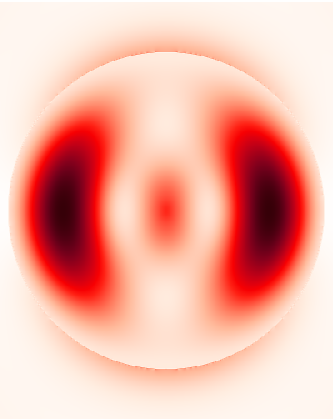}
\includegraphics[width=\subfigsize]{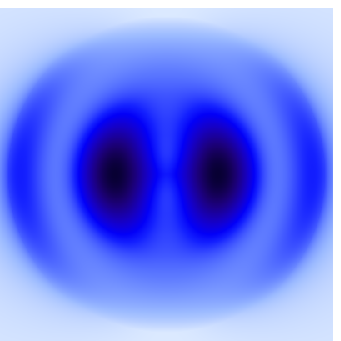}\\
{\centering
$r=190$~nm
\par}
\par} 
\caption{\label{ener_dist_anap} (Color online)
Time-average distributions of electric $|E|^2$ (left parts, red color) and magnetic $|H|^2$ (right parts, blue color) energy densities in the standalone GaP disks at resonances.
The distributions are calculated within the plane intersecting geometric center of the disk and parallel to its base.
The incident Gaussian beam is polarized along the vertical direction.}
\end{figure} 

The sizes of disks with Mie-type resonances at the wavelength of 532~nm should be identified.
This may be done by studying the field profiles.
Figure \ref{ener_dist_anap} illustrates the time-average electric $|E|^2$ or magnetic $|H|^2$ energy distributions inside standalone nanodisks at the sizes close to Mie-type resonances. 
In Fig.~\ref{ener_dist_anap}, the maximum values of energy density are 807 ($|E|^2$) and 239 ($|H|^2$) for $r=50$~nm, 1493 ($|E|^2$) and 642 ($|H|^2$) for $r=92$~nm, 6406 ($|E|^2$) and 4310 ($|H|^2$) for $r=126$~nm, 1203 ($|E|^2$) and 877 ($|H|^2$) for $r=190$~nm with respect to the beam intensity at the waist center.
The dipole magnetic resonance is observed near $r=50$~nm, the electric dipole resonance can be identified about $r=92$~nm. 

\begin{figure}
{\centering
\includegraphics[width=\figsize]{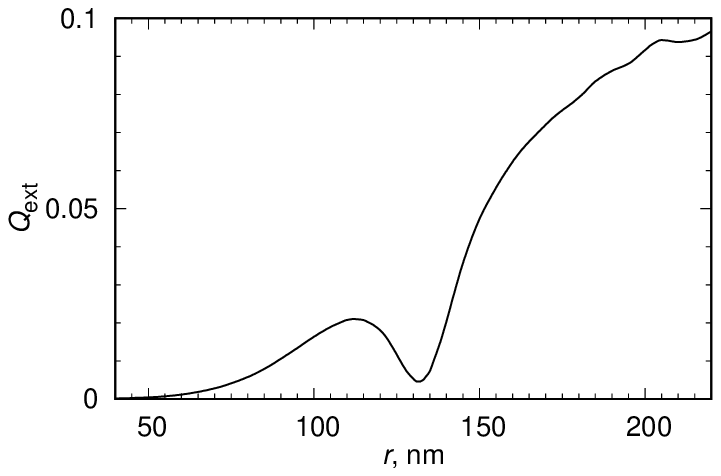}
\par} 
\caption{\label{qext_r_l532}
The extinction efficiency $Q_{\mathrm{ext}}$ of the standalone GaP nanodisk of thickness 50~nm versus its radius~$r$ at $\lambda=532$~nm. Light
incidence is normal to the top surface of the disk.}
\end{figure} 

The field profiles for the strong resonance just about $r=125$~nm are similar to those calculated for anapole states in Si nanodisks \cite{Miroshnichenko15,Baryshnikova19}.
The optical parameters used for Si nanodisk modeling ($n=4$, $\lambda=550$~nm, $r\approx 100$~nm and $h=50$~nm) are close to properties of the GaP particles at this resonance.
Fig.~\ref{qext_r_l532} shows a dip of extinction efficiency $Q_{\mathrm{ext}}$ of the standalone GaP nanodisks with $r=132$~nm at $\lambda=532$~nm.
This is due to the scattering suppression at the anapole mode.
This mode also exhibits the highest field enhancement inside the particle.
The size of the local minimum of $Q_{\mathrm{ext}}$ is slightly higher than that of maximum enhancement of the optical Kerr effect.
The similar phenomenon was observed both by numeric simulations and experimentally for maxima of Raman signal from silicon nanodisks at the anapole state \cite{Baranov18}.
This difference may be associated with the suppressed extinction of the particle at the anapole state and, as a consequence, the nonlinear effects confine within a disk.
Since FDTD simulations use dimensionless units and the ratio of sizes to the wavelength is constant, on the spectral dependence, the maximum nonlinear effect will be observed for a light frequency slightly higher than that of the anapole state.
Therefore, the GaP particle mode at this size can be recognized as the lowest order anapole state.

The time-average distribution of electric $|E|^2$ energy density for disk with $r=190$~nm is similar to the field profiles obtained for germanium nanodisks and denoted as a high-order mode (HOM) in Ref.~\cite{Grinblat17a} or ED$_2$ in Ref.~\cite{Grinblat16}. 
The extinction efficiency is enhanced for the disk with $r=190$~nm (Fig.~\ref{qext_r_l532}).
Therefore, this is not an anapole mode.
Nevertheless, there is also a transition to negative values of $n_{2\,\mathrm{eff}}(r)$ at this resonance.

Both dipole resonances do not reveal themselves on the dependencies $n_{2\,\mathrm{eff}}(r)$ (Figs.~\ref{n2_r_GaP_30_randcyl}, \ref{n2_r_GaP_36_lattcyl}).
As shown in Ref.~\cite{vandeGroep13}, for the magnetic and electric dipole modes in thin Si cylinders, the large fractions of current loops lie outside the particle.
Thus, there is no large field enhancement within the particle and the nonlinear effects are not magnified.
In contrast, the anapole state results in the giant enhancement of the effective second-order nonlinear refractive index and the inversion of its sign.
It should be emphasized that $|n_{2\,\mathrm{eff}}(r)|$ at the anapole state exceeds that of the metasurface consisting of GaP nanospheres at the dipole magnetic resonance 
\cite{Panov19}.
In Ref.~\cite{Panov19}, the anapole state for spheres was not observed while according to theory~\cite{Lukyanchuk17} it is expected at $r=115$~nm for GaP at $\lambda=532$~nm.
The excitation of the anapole mode in spherical particles with a plane wave is complicated \cite{Baryshnikova19}.
On the other hand, the anapole states dominate in disk particles as other modes are suppressed.

The resonance at $r=195$~nm also exhibit the inversion of $n_{2\,\mathrm{eff}}(r)$ sign.
The values of $|n_{2\,\mathrm{eff}}(r)|$ are below those at the anapole state by an order of magnitude and close to ones observed at the Mie resonances in the disordered metasurfaces consisting of the spheres \cite{Panov19}.

The comparison with metasurfaces consisting of spherical particles \cite{Panov19} shows that nanodisks at the anapole mode possess the higher optical nonlinearity.
The nanospheres at the first Mie resonances have smaller sizes than disks at the anapole state.
Nevertheless, the fabrication of the nanodisks on substrate is simpler.
The FDTD modeling shows a minor influence of the substrate with $n=1.5$ on the field profiles for GaP nanodisks. 

In conclusion, the second order refractive index of disordered  and square lattice metasurfaces of GaP nanodisks in the vicinity of the anapole state grows exponentially. 
At the magnetic and electric dipole resonances of the disks, the effective second order refractive index does not show abrupt jumps.
The sign of nonlinear Kerr coefficient is inverted at the anapole state as well as at non-anapole higher order resonance. 
The absolute value of the second order refractive index at the anapole mode by an order of magnitude exceeds that of the non-anapole resonance of the arrays of nanodisks or nanospheres.

\subsubsection*{Acknowledgements} 
 The results were obtained with the use of IACP FEB RAS Shared Resource Center ``Far Eastern Computing Resource'' equipment (https://www.cc.dvo.ru).

\subsubsection*{Disclosures} 
The author declares no conflicts of interest.

\bibliography{nlphase}


\end{document}